\documentclass[a4paper,11pt]{article}
\usepackage{pos}
\usepackage[utf8]{inputenc}
\usepackage{ragged2e}
\usepackage{calligra,amsmath,amsfonts,bbm,mathrsfs,amssymb}
\usepackage{slashed,cancel,units}
\usepackage{catchfile}
\usepackage{indentfirst}
\usepackage{graphicx}
\usepackage{float}
\usepackage{enumerate}
\usepackage{subcaption}
\usepackage{setspace}

\newcommand{\be}{\begin{equation}}
\newcommand{\ee}{\end{equation}}
\newcommand{\ba} {\begin{equation}\begin{aligned}}
\newcommand{\ea} {\end{aligned}\end{equation}}
\newcommand{\bea}{\begin{eqnarray}}
\newcommand{\eea}{\end{eqnarray}}

\def\diag{{\tt diag}}

\def\hc{\mathrm{h.c.}}

\newcommand{\VEV}[1]{\langle #1 \rangle}

\DeclareMathOperator{\unity}{\mathbbm{1}}
\def\TeV{\text{ TeV}}

\newcommand{\sL}{\mathscr{L}}
\newcommand{\cG}{\mathcal{G}}

\newcommand\sO{\mathscr{O}}

\newcommand\cy{\mathbf{y}}
\newcommand\cY{\mathcal{Y}}
\newcommand\cDY{\Delta\mathcal{Y}}
\newcommand\DY{\Delta Y}
\newcommand\cV{\mathcal{V}}

\title{The Data Driven Flavour Model}

\author{Luca Merlo}

\affiliation{Instituto de F\'isica Te\'orica UAM/CSIC and Departamento  de  F\'{\i}sica Te\'{o}rica,\\  
Universidad  Aut\'{o}noma  de  Madrid, Calle Nicol\'as Cabrera 13-15, Cantoblanco E-28049 Madrid, Spain}
 
\emailAdd{luca.merlo@uam.es}

\abstract{A bottom-up approach has been adopted to identify a flavour model that agrees with present experimental measurements. The charged fermion mass hierarchies suggest that only the top Yukawa term should be present at the renormalisable level. The flavour symmetry of the Lagrangian including the fermionic kinetic terms and only the top Yukawa is then a combination of $U(2)$ and $U(3)$ factors. Lighter charged fermion and active neutrino masses and quark and lepton mixings arise considering specific spurion fields. The associated phenomenology is investigated and the model turns out to have almost the same flavour protection of the Minimal Flavour Violation, in both quark and lepton sectors. Promoting the spurions to be dynamical fields, the associated scalar potential is also studied and a minimum is identified such that fermion masses and mixings are correctly reproduced.}

\FullConference{%
  40th International Conference on High Energy physics - ICHEP2020\\
  July 28 - August 6, 2020\\
  Prague, Czech Republic (virtual meeting)
}

\begin{document}
\maketitle

\section{Introduction}

The seek of an explanation for the heterogeneity of fermion masses and mixings is nowadays one of the biggest issues in particle physics. The success of the Standard Model (SM) in describing the strong and electroweak interactions through a gauged symmetry encourages the idea that flavour symmetries may provide a solution to this problem. There are many examples in the literature, using Abelian and non-Abelian, discrete and continuous, global or local symmetries~\cite{Froggatt:1978nt,Altarelli:2000fu,Altarelli:2012ia,Bergstrom:2014owa,Ma:2001dn,Altarelli:2005yp,Altarelli:2005yx,Feruglio:2007uu,Feruglio:2008ht,Bazzocchi:2009pv,Bazzocchi:2009da,Altarelli:2009gn,AristizabalSierra:2009ex,Feruglio:2009iu,Lin:2009sq,Feruglio:2009hu,Toorop:2010yh,Varzielas:2010mp,Altarelli:2010gt,Toorop:2010ex,Toorop:2010kt,Merlo:2011hw,Altarelli:2012bn,Altarelli:2012ss,Bazzocchi:2012st,DAmbrosio:2002vsn,Cirigliano:2005ck,Davidson:2006bd,Grinstein:2010ve,Alonso:2011yg,Alonso:2011jd,Barbieri:2011ci,Buras:2011zb,Buras:2011wi,Alonso:2012jc,Alonso:2012fy,Alonso:2012pz,Blankenburg:2012nx,Lopez-Honorez:2013wla,Alonso:2013mca,Alonso:2016onw,Dinh:2017smk,Arias-Aragon:2017eww,Merlo:2018rin,Arias-Aragon:2020bzy,Arias-Aragon:2020qip}. 

The Data Driven Flavour Model (DDFV)~\cite{Arias-Aragon:2020bzy} is a bottom-up approach based on a continuous global symmetry, where the main idea is to strictly follow what data suggests, avoiding any additional requirement for a specular treatment of all the fermions species: within the SM context with or without the addition of three RH neutrinos, the criterium is that only the term corresponding to the top quark mass and, if existing, to the RH neutrino Majorana masses are invariant under the considered flavour symmetry without any spurion insertion, while the Yukawa terms for the other fermions need these insertions. The schematic structure for the resulting Yukawa and mass matrices, writing the Lagrangian in the left-right notation, looks like
\be
\begin{gathered}
Y_U=\left(
\begin{array}{cc|c}
x & x & 0 \\
x & x & 0 \\
\hline
0 & 0 & 1 \\
\end{array}\right)\,,\qquad\qquad
Y_D=\left(
\begin{array}{ccc}
x & x & x \\
x & x & x \\
\hline
y & y & y \\
\end{array}\right)\,,\\
m_\nu\propto\left(
\begin{array}{ccc}
x & x & x \\
x & x & x \\
x & x & x \\
\end{array}\right)\,,\qquad\qquad
Y_E=\left(
\begin{array}{cc|c}
x & x & y \\
x & x & y \\
x & x & y \\
\end{array}\right)\,,
\end{gathered}
\ee
where $m_\nu$ is the neutrino mass matrix as arises from the Weinberg operator. The $x$ and $y$ entries represent spurion background contributions and are numbers smaller than 1. The vertical and horizontal lines help identifying the flavour structures. 

For the type I Seesaw case, the neutrino sector is instead described by a Dirac Yukawa matrix and a Majorana mass matrix as follows:
\be
\begin{gathered}
Y_\nu=\left(
\begin{array}{ccc}
x & x & x \\
x & x & x \\
x & x & x \\
\end{array}\right)\,,\qquad\qquad
M_R\propto\unity\,.
\end{gathered}
\ee

The advantages of this model are multiple: it distinguishes the third families from the lighter ones; it naturally describes the top Yukawa of order 1, avoiding any technical difficulty for the perturbative expansion in the case of promoting spurions to flavons; it explains the smallness of the bottom and tau masses with respect to the top mass without any additional assumption; it assigns neutrinos to the same flavour representation, as suggested by the largeness of the atmospheric and solar mixing angles. This model is therefore a bottom-up approach, completely data driven, that encodes the advantages of the Minimal Flavour Violation (MFV) approach~\cite{Chivukula:1987py,DAmbrosio:2002vsn,Cirigliano:2005ck,Davidson:2006bd,Gavela:2009cd,Alonso:2011jd} and of the $U(2)^n$ model~\cite{Barbieri:2011ci,Blankenburg:2012nx}, avoiding their major drawbacks.

\section{The Model}

The Lagrangian of the DDFM model can be written as the sum of different terms,
\be
\sL=\sL_\text{kin}+\sL_\text{Y}-\cV(\phi)\,,
\ee
where $\sL_\text{kin}$ contains the canonical kinetic terms of all the fields in the spectrum, $\cV(\phi)$ stands for the SM scalar potential of the Higgs doublet $\phi$, and $\sL_\text{Y}$ is responsible for the fermion masses. \\

\noindent{\bf Quark Sector}

The $\sL_\text{Y}$ part of the Lagrangian for the quark sector  can be written as
\be
-\sL^q_\text{Y}=y_t\,\bar{q}'_{3L}\,\tilde\phi\, t'_R + \Delta\sL^q_\text{Y}+\hc\,,
\ee
where $q'_{3L}$ stands for the $SU(2)_L$ doublet of the left-handed (LH) third family quarks, $t'_R$ for the $SU(2)_L$ singlet RH top quark, $\tilde\phi=i\sigma_2\phi^*$, and $\Delta\sL^q_\text{Y}$ contains all the terms responsible for the other quark masses and quark mixings. The prime identifies the flavour or interaction basis. 
The largest non-Abelian quark flavour symmetry consistent with the whole Lagrangian, neglecting $\Delta\sL^q_\text{Y}$, is given by
\be
\cG_q=SU(2)_{q_L}\times SU(2)_{u_R}\times SU(3)_{d_R}\,,
\ee
where the notation matches the one of MFV as seen in the introduction. The fields $q'_{3L}$ and $t'_R$ appearing in $\sL^q_\text{Y}$ are singlets under $\cG_q$. The other quark fields, instead, transform non-trivially: the LH quarks of the first two families, labelled as $Q'_L$, transform as a doublet under $SU(2)_{q_L}$; the RH up-type quarks of the first two families, indicated by $U'_R$, transform as a doublet under $SU(2)_{u_R}$; finally, the three RH down-type quarks, $D'_R$, transform altogether as a triplet of $SU(3)_{d_R}$. 

The lighter families and the mixing are described in $\Delta\sL^q_\text{Y}$, once a specific set of spurions are considered. In order to keep the model as minimal as possible, only three spurions are introduced: $\cDY_U$ that is a bi-doublet of $SU(2)_{q_L}\times SU(2)_{u_R}$, $\cDY_D$ that is a doublet-triplet of $SU(2)_{q_L}\times SU(3)_{d_R}$, and $\cy_D$ that is a vector triplet of $SU(3)_{d_R}$. These spurions develop background values such that the resulting Yukawa matrices are then given by
\be
Y_U=\left(
\begin{array}{cc}
\VEV{\cDY_U} & 0 \\
0 & 1 \\
\end{array}\right)\,,\qquad\qquad
Y_D=\left(
\begin{array}{c}
\VEV{\DY_D} \\
\VEV{y_D} \\
\end{array}\right)\,,
\label{Yukawas}
\ee
where the $Y_U$ is already diagonal, while $Y_D$ is exactly diagonalised by the CKM matrix.\\

\noindent{\bf Lepton Sector: Extended Field Content (EFC)}

When considering the type I Seesaw context, three RH neutrinos are added to the SM spectrum and their masses are assumed to be much larger than the electroweak scale. It follows that the lepton Yukawa Lagrangian can be written as
\be
-\sL^{\ell,\text{EFC}}_\text{Y}=\dfrac{1}{2}\Lambda_{LN}\,\bar{N}^{\prime c}_R\,Y_N\,N'_R+\Delta\sL^{\ell,\text{EFC}}_\text{Y}+\hc\,,
\ee
where $\Lambda_{LN}$ is an overall scale associated to lepton number violation, $Y_N$ is a dimensionless matrix, and $\Delta\sL^{\ell,\text{EFC}}_\text{Y}$ contains all the terms responsible for the other lepton masses and mixing. If $Y_N$ is a completely generic matrix, the model turns out to be non-interesting. For this reason $Y_N$ is taken to be the identity matrix. In this special case, the lepton flavour symmetry is given by 
\be
\cG^\text{EFC}_\ell=SU(3)_{\ell_L}\times SU(2)_{e_R}\times SO(3)_{N_R}\,,
\ee
where the LH doublets transform as a triplet of $SU(3)_{\ell_L}$, the RH charged leptons as a doublet$+$singlet of $SU(2)_{e_R}$ and the RH neutrinos transform as a triplet of $SO(3)_{N_R}$. An interesting aspect of this choice is that it is compatible with the $SU(5)$ grand unification setup, that may be an ultraviolet completion of the model presented here. 

Lepton masses and mixing are described by means of three spurions: $\cDY_E$ that transforms as a triplet-doublet of $SU(3)_{\ell_L}\times SU(2)_{e_R}$, $\cy_E$ as a vector triplet of $SU(3)_{\ell_L}$, and finally $\cY_\nu$ as a bi-triplet under $SU(3)_{\ell_L}\times SO(3)_{N_R}$. Once the spurions acquire precise background values, 
\be
Y_E=\left(
\begin{array}{cc}
\langle\cDY_E\rangle & \langle\cy_E\rangle \\
\end{array}\right)\,,\qquad\qquad
\diag(m_{\nu_1},\,m_{\nu_2},\,m_{\nu_3})=U^T\,
\dfrac{v^2}{2\Lambda_{LN}}\,\langle \cY_\nu^*\rangle \langle \cY_\nu^\dag\rangle
\, U\,,
\label{YEtotal}
\ee
the charged lepton mass matrix is already diagonal, while the neutrino mass matrix can be diagonalised by the PMNS matrix.

\section{Phenomenological Analysis}

The analysis is carried out adopting an effective field theory approach and considering operators with at most mass dimension six. 

In the quark sector, the bounds on the dimension 6 operators within the DDFM are the same as in the MFV framework and representative examples are reported in Tab.~\ref{TabBounds}~\cite{Isidori:2012ts}.

\begin{table}[h!]
\begin{center}
\begin{tabular}{|c|c|c|}
\hline
&&\\[-3mm]
Operators & Bound on $\Lambda/\sqrt{a_i}$ & Observables \\[1mm]
\hline
&&\\[-3mm]
$\sO_{1}$, $\sO_{2}$ & $5.9\TeV$ & $\epsilon_K$, $\Delta m_{B_d}$. $\Delta m_{B_s}$ \\[1mm]
$\sO_{17}$, $\sO_{18}$ & $4.1\TeV$ & $B_s\to\mu^+\mu^-$, $B\to K^\ast\mu^+\mu^-$ \\[1mm]
$\sO_{21}$, $\sO_{22}$ & $3.4\TeV$ & $B\to X_s\gamma$, $B\to X_s \ell^+\ell^-$ \\[1mm]
$\sO_{25}$, $\sO_{26}$ & $6.1\TeV$ & $B\to X_s\gamma$, $B\to X_s \ell^+\ell^-$ \\[1mm]
$\sO_{27}$, $\sO_{28}$ & $1.7\TeV$ & $B\to K^\ast\mu^+\mu^-$ \\[1mm]
$\sO_{29}$, $\sO_{30}$, $\sO_{31}$, $\sO_{32}$ & $5.7\TeV$ & $B_s\to\mu^+\mu^-$, $B\to K^\ast\mu^+\mu^-$ \\[1mm]
$\sO_{33}$, $\sO_{34}$ & $5.7\TeV$ & $B_s\to\mu^+\mu^-$, $B\to K^\ast\mu^+\mu^-$ \\[1mm]
\hline
\end{tabular}
\end{center}
\caption{\em Lower bounds on the NP scale for some representative effective dimension 6 operators. The values of $\Lambda$ are at $95\%$ C.L. and are obtained considering that only the operators of the same class contribute to the given observables. The labelling of the operators corresponds to the original paper.}
\label{TabBounds}
\end{table}%

The bounds turn out to be in the TeV range and this suggests that precision investigations in rare decays together with complementary studies at colliders may play a key role to unveil the physics behind the flavour sector. The only difference between MFV and DDFV is in the presence of some decorrelations associated to the charged leptons: the decay rates for $B_s\to\mu^+\mu^-$  and $B_s\to\tau^+\tau^-$ are predicted to be exactly the same as in MFV, while they are independent observables in the DDFM; similarly for $B\to K^\ast\mu^+\mu^-$ and $B\to K^\ast\tau^+\tau^-$.

In the lepton sector, the results are very similar to MLFV, but with small differences due to the decorrelation of observables associated to the tau. These effects may be seen explicitly in ratios of branching ratios of rare radiative decays. \\

Promoting the spurions to  dynamical fields gives the possibility to shed some light on the possible dynamical origin of the flavour structures responsible for the phenomenological results of the DDFM. The analysis reveals that a minimum exists where all the masses and mixings can indeed be described in agreement with data, but at the price of tuning some parameters of the scalar potential. Moreover, precise predictions for the leptonic Dirac and Majorana phases follow from the minimisation of the scalar potential: this is a difference with the MLFV framework, where strictly CP conserving phases are allowed and it results in very different predictions for the neutrinoless-double-beta decay.

Although this may not be considered the ultimate solution to the flavour puzzle, it represents a step ahead to achieve this goal and a significant improvement with respect to MFV, where only part of masses and mixing can be correctly described. 

\section*{Acknowledgements}

The author acknowledges partial financial support by the Spanish MINECO through the Centro de excelencia Severo Ochoa Program under grant SEV-2016-0597, by the Spanish ``Agencia Estatal de Investigac\'ion''(AEI) and the EU ``Fondo Europeo de Desarrollo Regional'' (FEDER) through the projects FPA2016-78645-P and PID2019-108892RB-I00/AEI/10.13039/501100011033, and by the Spanish MINECO through the ``Ram\'on y Cajal'' programme (RYC-2015-17173).


\footnotesize
{\setstretch{0.5}

\providecommand{\href}[2]{#2}\begingroup\raggedright\endgroup

}
\end{document}